# The Effects of Geometry on a-Si:H Solar Cell Performance

T. Kirkpatrick[*], M. J. Burns, M. J. Naughton

Department of Physics, Boston College, 140 Commonwealth Avenue, Chestnut Hill, Massachusetts 02467, USA

[*]Present Address: Department of Mechanical Engineering, Massachusetts Institute of Technology, 77 Massachusetts Avenue, Cambridge, Massachusetts 02139, USA

Corresponding author: Timothy Kirkpatrick, kirktim@mit.edu

**ABSTRACT**

We present a model for simulating performance of 3D nano -coaxial and -hemispherical thin film solar cells. The material system considered in these simulations is hydrogenated amorphous silicon (a-Si:H), with solar cells fabricated in an *n-i-p* stacking architecture. Simulations for the performance of the planar a-Si:H device are compared against simulations performed using SCAPS-1D and found to be in close agreement. Electrical and optical properties of devices are discussed for the respective geometries. Maximum power point efficiencies are plotted as a function of *i*-layer thickness for insight into optimizing spatial parameters. Simulation results show that while geometrical changes in the energy band diagram impact charge carrier collection, a-Si:H solar cell performance is most significantly impacted by light absorption properties associated with nanoscopic arrays of non-planar structures. We compare our simulations to results of fabricated nanocoaxial a-Si:H solar cells and infer the mechanisms of enhanced absorption observed experimentally in such solar cells.

## 1. Section I (Introduction)

Because single-crystal semiconductors can be expensive to produce [1-4], there has been a great deal of interest in producing solar cells from inexpensive thin film techniques [5-12], utilizing physical and/or chemical deposition methods. In addition to being comparatively inexpensive to produce, amorphous semiconductor absorption coefficients tend to be significantly larger than those of crystalline semiconductors across the majority of the visible spectrum [3, 6, 10-17], reducing the volume of material necessary to capture incident light, thereby further reducing the overall cost of solar cell fabrication. However, compared to crystalline materials, amorphous materials have little or no long-range atomic order and, in addition, often contain intrinsic defects, which tend to increase the density of trap states within the band gap. This has a compounded detrimental affect on the electrical properties [18-22] of amorphous materials by decreasing charge carrier mobility $\mu_\nu$, and decreasing charge carrier lifetime $\tau_\nu$. The product of these two parameters is crucial in determining the diffusion and drift lengths of charge carriers in semiconducting materials. Low values for $\mu_\nu$ and $\tau_\nu$, in turn, yield larger amounts of dark-current for amorphous solar cell devices.

One design criterion for maximizing efficiency of a-Si:H solar cells (as well as crystalline cells to a lesser extent) stems from the orientation of the photovoltaic junction with which the devices are fabricated; *i.e.* in a planar geometry. Because optical and electronic path lengths are, on average, collinear in planar geometries, the largest possible solar cell efficiencies occur for materials with electronic diffusion/drift lengths that are much greater than average photon absorption depths [23]. Despite an enhancement in absorption over crystalline counterparts, amorphous materials do not fit this criterion [3, 5-14], which is indicative of just how low the $\mu_\nu$ and $\tau_\nu$ values are. However, non-planar solar cell geometries do not have collinear electronic and optical path lengths. Therefore, it is possible that by orthogonalizing these two path lengths using a non-planar architecture, solar cell efficiency may improve, despite low $\mu_\nu$ and $\tau_\nu$ values, by creating devices which are electrically "thin" in one direction and optically "thick" in another [24]. Previous work establishing a formal mathematical framework for analytically modeling geometrically generalized non-planar solar cells showed that device geometry and material properties are inextricably linked to overall solar cell performance [23]. Results qualitatively agreed with physical arguments about mutually orthogonal electronic and optical path lengths and, in addition, quantitatively showed that one design, in particular, significantly improved efficiency when using materials with properties that induce short electronic path lengths (*i.e.* low $\mu_\nu$ and $\tau_\nu$) with respect to average absorption depths. For materials where the average absorption depth was smaller than, and even on the order of, the electronic diffusion/drift lengths, little improvement in efficiency was observed for non-planar geometries over the planar geometry [23]. Therefore, amorphous materials represent an ideal material system to perform more detailed simulations in non-planar solar cell architectures. Here, we simulate performance of a-Si:H *n-i-p* solar cells for comparison in planar, coaxial, and hemispherical designs (see Fig. 1).

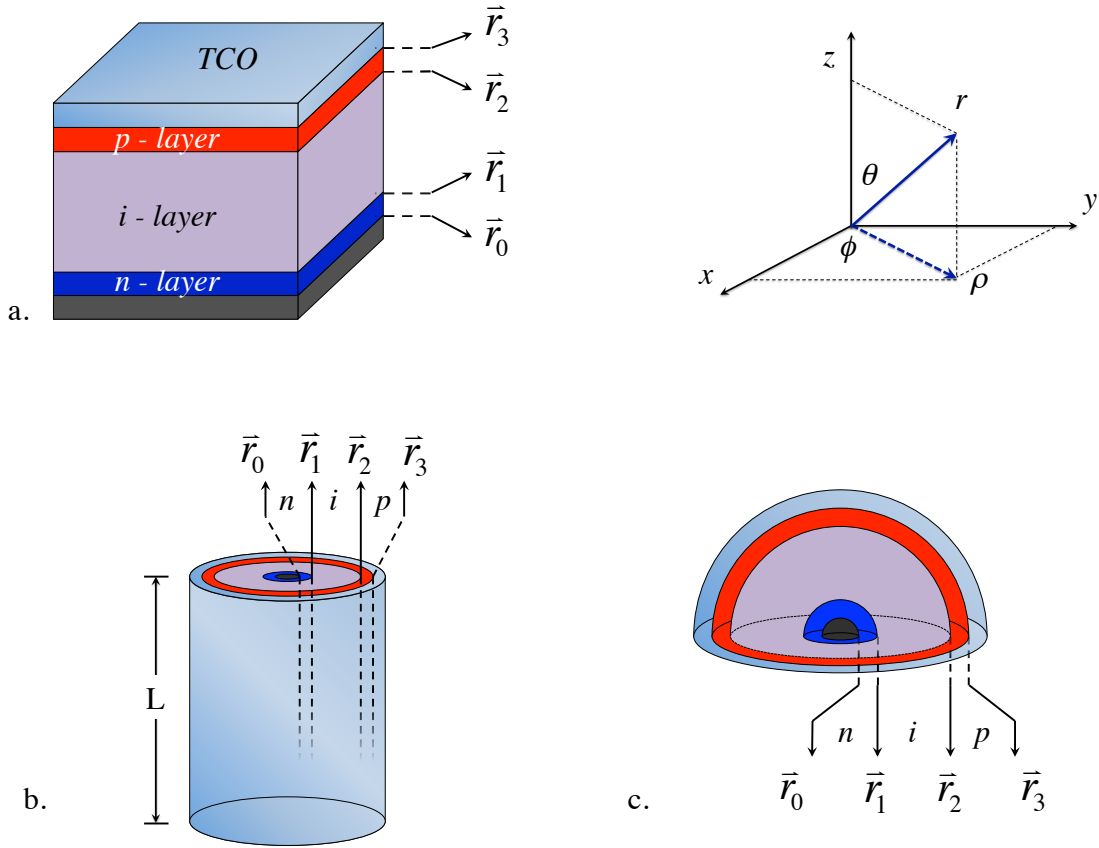

**Fig. 1**. Architectures considered for a-Si:H, *n-i-p* solar cells; a) planar, b) coaxial, and c) hemispherical configurations.

## 2. Section II (Theory)

Our model [23] describing a generalized framework for analytically modeling non-planar solar cell architectures emphasized recombination variability in the space-charge region (SCR) (where the majority of the energy band bending occurs) as a function of geometry [23]. The *i*-layer thickness in a-Si:H solar cells is typically an order of magnitude larger than both the *p*- and *n*-layers, thereby dominating overall solar cell performance [3-5, 12]. Because the middle layer is intrinsic (*i.e.* it has a much lower charge carrier concentration), the *p*- and *n*- layers create a large depletion region in the *i*-layer relative to the *i*-layer thickness. For typical *i*-layer thicknesses used in a-Si:H solar cells (~ 100 nm), the entire *i*-layer is essentially all space-charge [4]. As such, this model is well suited to study device performance of non-planar a-Si:H solar cells.

Because of relative layer thicknesses, our simulations for a-Si solar cells only emphasize the *i*-layer, and neglect any diffusional transport from the *p*- and *n*-doped quasi-neutral regions. As a validity check in making this approximation, in Fig. 2 we compare device performance simulations for a *planar* a-Si:H solar cell with, and without, electrical contributions from the *p*- and *n*-layers. Results from the planar architecture indicate that device performance of a-Si:H is negligibly impacted by quasi-neutral region transport, provided the *i*-layer is much thicker than the *p*- and *n*- layers. In these simulations, the ratios of *n*:*i*:*p* layer thicknesses are held constant at 5:60:6 nm, and *i*-layer thickness is used as a batching parameter in the efficiency calculation of Fig. 2. In addition, we also compare our simulations for a planar a-Si:H *n*-*i*-*p* solar cell against a standard in amorphous material solar cell simulations, SCAPS-1D [25]. Using the same material parameters, SCAPS returns a similar efficiency curve to ours, with the primary difference being that efficiency values are slightly higher than ours across the breadth of *i*-layer thickness values (see Fig. 2). For these simulations, the peak efficiency occurs near an *i*-layer thickness of ~ 200 nm, with SCAPS predicting an efficiency approximately 1.5% (absolute) higher than that predicted using our simulation. It should be noted that SCAPS takes into account band-tail states, while our model does not. Both utilize mid-gap trap states in the *i*-layer, approximated to be the same energy level as the intrinsic chemical potential.

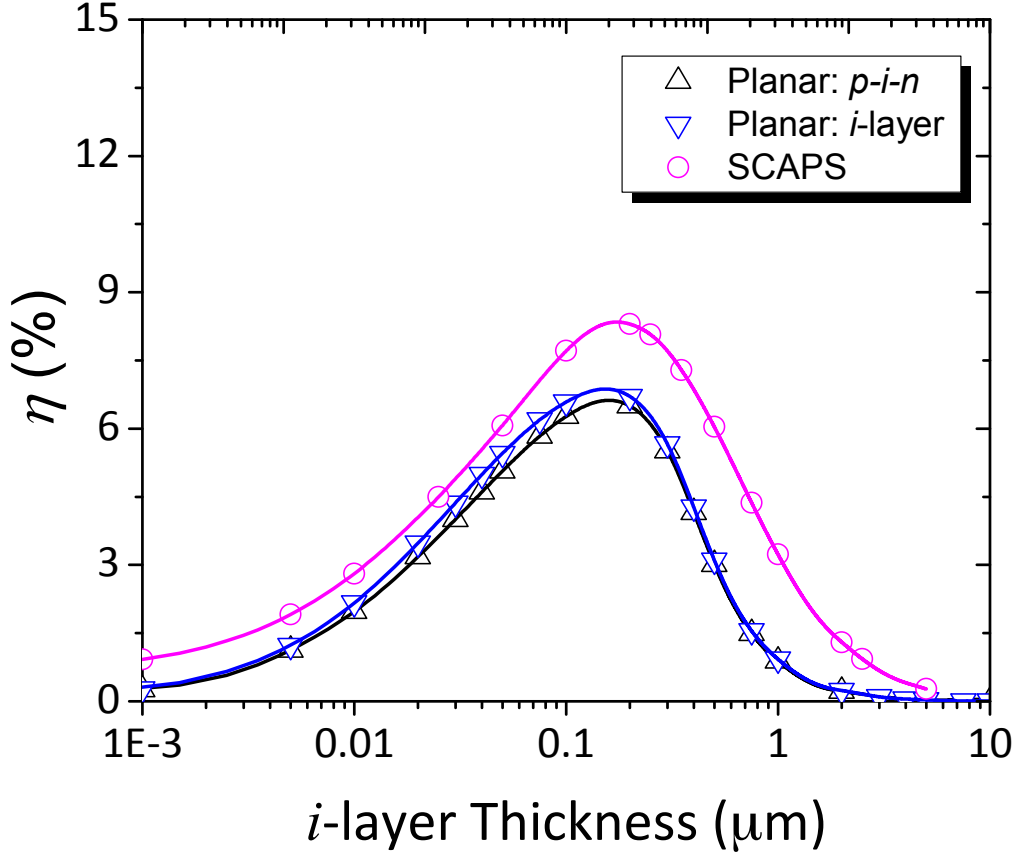

**Fig. 2**. Efficiency curves for planar a-Si:H solar cells as a function of *i*-layer thickness. The results indicate that total device performance of planar a-Si:H solar cells is negligibly impacted by quasi-neutral region transport, provided that the *i*-layer is much thicker than the *p*- and *n*- layers. For these simulations, the ratio of *n:i:p* layer thicknesses is held constant at 5:60:6 nm.

Unlike the planar geometry, where current conservation together with constant device cross section results in conservation of current *density*, in non-planar geometries, the non-constant device cross section results in only current being conserved. For any geometrical orientation of a photovoltaic junction aligned symmetrically along a single axis, the derived [23] contribution from the SCR to the total current of the device is given by the spatial integral of the generation and recombination rates over the SCR volume; *i.e.*

$$i_{SC} = q \iiint [G_{SC}(\vec{r}) - U_{SC}(\vec{r})] d^3 r.$$

For device performance simulations of a-Si:H solar cells, we again note that this equation is appropriate because performance is dominated by an assumed fully depleted *i*-layer, which is at least an order of magnitude thicker than the *p*- and *n*-layers. However, to more accurately account for imperfect charge carrier collection in the device, we insert an ad hoc charge carrier collection probability factor $\psi(\vec{r})$ with the generation current,

$$i \approx i_{SC} = q \iiint [G_{SC}(\vec{r})\psi(\vec{r}) - U_{SC}(\vec{r})]d^3r.$$

For all solar cell configurations, we take light to be entering through the *p*-type window and/or along the z-axis. Solar cell size $A_{PV}$ is set to $1\ cm^2$ for all devices. Material properties and simulation parameters are listed in the Appendix. Expressions for a-Si:H solar cell current in planar, coaxial, and hemispherical architectures are discussed in detail in [23], and are reproduced here in Table 1.

**Table 1**. Expressions for a-Si:H current for planar, coaxial, and hemispherical structures.

| Geometry | Total Current Expression |
|---|---|
| Planar | $i = qA_{PV} \int_{r_1}^{r_2} [G_{SC}(z)\psi(z) - U_{SC}(z)]\, dz$ |
| Coaxial | $i = qN \iiint_{r_1}^{r_2} [G_{SC}(z)\psi(\rho) - U_{SC}(\rho)]\, \rho d\rho d\phi dz$ |
| Hemispherical | $i = qN \iiint_{r_1}^{r_2} [G_{SC}(r,\theta)\psi(r) - U_{SC}(r)]\, r^2\sin(\theta)drd\theta d\phi$ |

The recombination rate $U_{SC}(\vec{r})$ used in these calculations is a sum of radiative, Shockley-Reade-Hall (SRH), and Auger recombination in the SCR, as explained in [23]. The spatial dependence for all recombination is expressed implicitly in the intrinsic chemical potential $\mu_i(\vec{r})$. The spatial dependence of $\mu_i(\vec{r})$ is also discussed in [23]. Functional expressions for the generation rates in the *i*-layer are also detailed in [23] and reproduced here in Table 2. The upper limit for integration in Table 2 is a function of applied bias $V$, given by the expression $\varepsilon_{max}(V) = \chi_{SC} + \Delta_{SC} - V$, where $\Delta_{SC}$ is the band gap and $\chi_{SC}$ is the electron affinity in the *i*-layer. To reduce computation time in our simulations, we approximate the upper limit in the integral calculations to be $\varepsilon_{max}(0) = \chi_{SC} + \Delta_{SC}$.

**Table 2**. Functional expressions for generation rates with longitudinal light incidence $\hat{k} = -\hat{z}$, and approximate electric fields in the SCR in planar, coaxial, and hemispherical geometries.

| Geometry | Generation rate: $G_{SC}(\vec{r})$ | Electric field: $E(\vec{r}, V)$ |
|---|---|---|
| Planar $\hat{k} = -\hat{z}$ | $\int_{\Delta_{SC}}^{\varepsilon_{max}(V)} \frac{I_{AM1.5}(\varepsilon_\gamma)}{\varepsilon_\gamma} \alpha(\varepsilon_\gamma) \exp(-\alpha(\varepsilon_\gamma)[r_3 - z])\, d\varepsilon_\gamma$ | $\dfrac{[V_{B.I.} - V]}{r_2 - r_1}$ |
| Coaxial $\hat{k} = -\hat{z}$ | $\int_{\Delta_{SC}}^{\varepsilon_{max}(V)} \frac{I_{AM1.5}(\varepsilon_\gamma)}{\varepsilon_\gamma} \alpha(\varepsilon_\gamma) \exp(-\alpha(\varepsilon_\gamma)[L - z])\, d\varepsilon_\gamma$ | $\dfrac{[V_{BI} - V]}{\rho} \ln\left(\dfrac{r_2}{r_1}\right)$ |

| Hemispherical $\hat{k}=-\hat{z}$ | $\cos^2(\theta) \int_{\Delta_{SC}}^{\varepsilon_{max}(V)} \frac{I_{AM1.5}(\varepsilon_\gamma)}{\varepsilon_\gamma} \alpha(\varepsilon_\gamma) \exp\left(-\alpha(\varepsilon_\gamma)[r_3-r]\cos(\theta)\right) d\varepsilon_\gamma$ | $\frac{[V_{BI}-V]}{r^2}\begin{bmatrix} r_1\, r_2 \\ r_2-r_1 \end{bmatrix}$ |
|---|---|---|

The most crucial component for approximating the total current of the a-Si:H solar cell from the current in the *i*-layer is the charge carrier collection probability $\psi(\vec{r})$. Because the *i*-layer does not contain majority or minority charge carriers, the probability of extraction for both electrons and holes must be accounted for with this collection probability term; *i.e.* $\psi(\vec{r}) = \psi_e(\vec{r}) + \psi_h(\vec{r})$. For this probability, we assume that individual charge carrier collection decays exponentially away from the region where electrons and holes are collected, respectively, modulated by the *drift* lengths $\vec{l}_{v=e,h}(\vec{r},V)$ in the *i*-layer. After charge carriers are photogenerated in the *i*-layer, electrons are collected at the *n*-layer contact, and holes at the *p*-layer contact. Because the generation rate describes generated electron-hole pairs, the extraction probability $\psi(\vec{r})$, used here in conjunction with the generation rate that defines the light-current produced, must account for the probability of extracting both charge carrier types. Therefore, the sum of electron and hole collection probabilities can never be greater than one; *i.e.*

$$\psi(\vec{r}) = \psi_e(\vec{r}) + \psi_h(\vec{r}) = A(V)\exp\left(-\frac{\vec{r}-\vec{r}_1}{\vec{l}_e(\vec{r},V)}\right) + B(V)\exp\left(\frac{\vec{r}-\vec{r}_2}{\vec{l}_h(\vec{r},V)}\right) \leq 1,$$

where $A(V)$ and $B(V)$ are to be determined below. Because electrons are collected at the *n*-layer contact, for an electron-hole pair photogenerated at $\vec{r}_1$, the probability that the electron will be collected must be one. We, therefore, approximate that the sum of the electron and hole collection probabilities at $\vec{r}_1$ be one,

$$\psi(\vec{r})|_{\vec{r}=\vec{r}_1} = \psi_e(\vec{r})|_{\vec{r}=\vec{r}_1} + \psi_h(\vec{r})|_{\vec{r}=\vec{r}_1} = A(V) + B(V)\exp\left(\frac{\vec{r}_1-\vec{r}_2}{\vec{l}_h(\vec{r}_1,V)}\right) = 1,$$

to account for the collection probability of both charge carrier types associated with the generated current in this model. It is important to emphasize that the recombination rate across the *i*-layer will account for dark-current produced in this model, and the collection probability only accounts for a reduction in the light-current produced from imperfect charge carrier extraction in the *i*-layer. Likewise, for an electron-hole pair photogenerated at $\vec{r}_2$, the probability that the hole will be collected must also be one, and we therefore approximate that

$$\psi(\vec{r},V)|_{\vec{r}=\vec{r}_2} = \psi_e(\vec{r},V)|_{\vec{r}=\vec{r}_2} + \psi_h(\vec{r},V)|_{\vec{r}=\vec{r}_2} = A(V)\exp\left(-\frac{\vec{r}_2-\vec{r}_1}{\vec{l}_e(\vec{r}_2,V)}\right) + B(V) = 1.$$

Solving for the coefficients, $A(V)$ and $B(V)$,

$$A(V) = \frac{1-\exp\left(\frac{\vec{r}_2-\vec{r}_1}{\vec{l}_h(\vec{r}_1,V)}\right)}{\exp\left(\frac{\vec{r}_1-\vec{r}_2}{\vec{l}_e(\vec{r}_2,V)}\right) - \exp\left(\frac{\vec{r}_2-\vec{r}_1}{\vec{l}_h(\vec{r}_1,V)}\right)},$$

and

$$B(V) = \frac{exp\left(\frac{\vec{r}_2 - \vec{r}_1}{\vec{l}_e(\vec{r}_2, V)}\right) - 1}{exp\left(\frac{\vec{r}_2 - \vec{r}_1}{\vec{l}_e(\vec{r}_2, V)}\right) - exp\left(\frac{\vec{r}_1 - \vec{r}_2}{\vec{l}_h(\vec{r}_1, V)}\right)}.$$

The charge carrier drift lengths $\vec{l}_{v=e,h}(\vec{r}, V)$ in the *i*-layer are expressed in terms of the electric field $E(\vec{r}, V)$ as

$$\vec{l}_v(\vec{r}, V) = \mu_v \tau_v E(\vec{r}, V).$$

Based on approximate energy band diagram profiles [23], the electric fields in the SCR for the planar, coaxial, and hemispherical geometries are given in Table 2. Details of the energy band profiles used for non-planar architectures are discussed in detail in [23]. From the spatially dependent behavior of the electric fields for the non-planar structures, it is seen that charge carrier drift lengths $\vec{l}_v(\vec{r}, V)$ will decay with respect to the inner-most charge collecting region which, in turn, will affect the charge collection probability for each charge carrier type.

### 3. Section III (Results and Discussion)

Fig. 3 shows how charge carrier collection begins to decrease with increasing *i*-layer thickness. The coaxial architecture collects charge more efficiently for thicker *i*-layers, as indicated in Fig. 3b and 3c, however, this result will change for varying inner radii $\vec{r}_0$ values used in the calculations; the initial electric field intensity at $\vec{r}_1$ will vary as $\vec{r}_0$ varies, for both the coaxial and hemispherical structures. The asymmetry in the planar charge carrier collection probability (the planar electric field is approximated as constant) arises from the difference in drift lengths for electrons and holes, due to the two-order of magnitude lower hole mobility associated with intrinsic a-Si:H [3, 5-9, 12-14, 18-22, 26]. Because of this, for thicker *i*-layers, the probability of hole collection is extremely low, except when electron-hole pairs are photogenerated very near the *p*-layer interface (see Fig. 3a). For the non-planar architectures, this asymmetry is compounded because, in addition to shorter hole drift-lengths, the electric field decays in intensity from $\vec{r}_1$ as $\rho^{-1}$ and $r^{-2}$ for the coaxial and hemispherical structures, respectively.

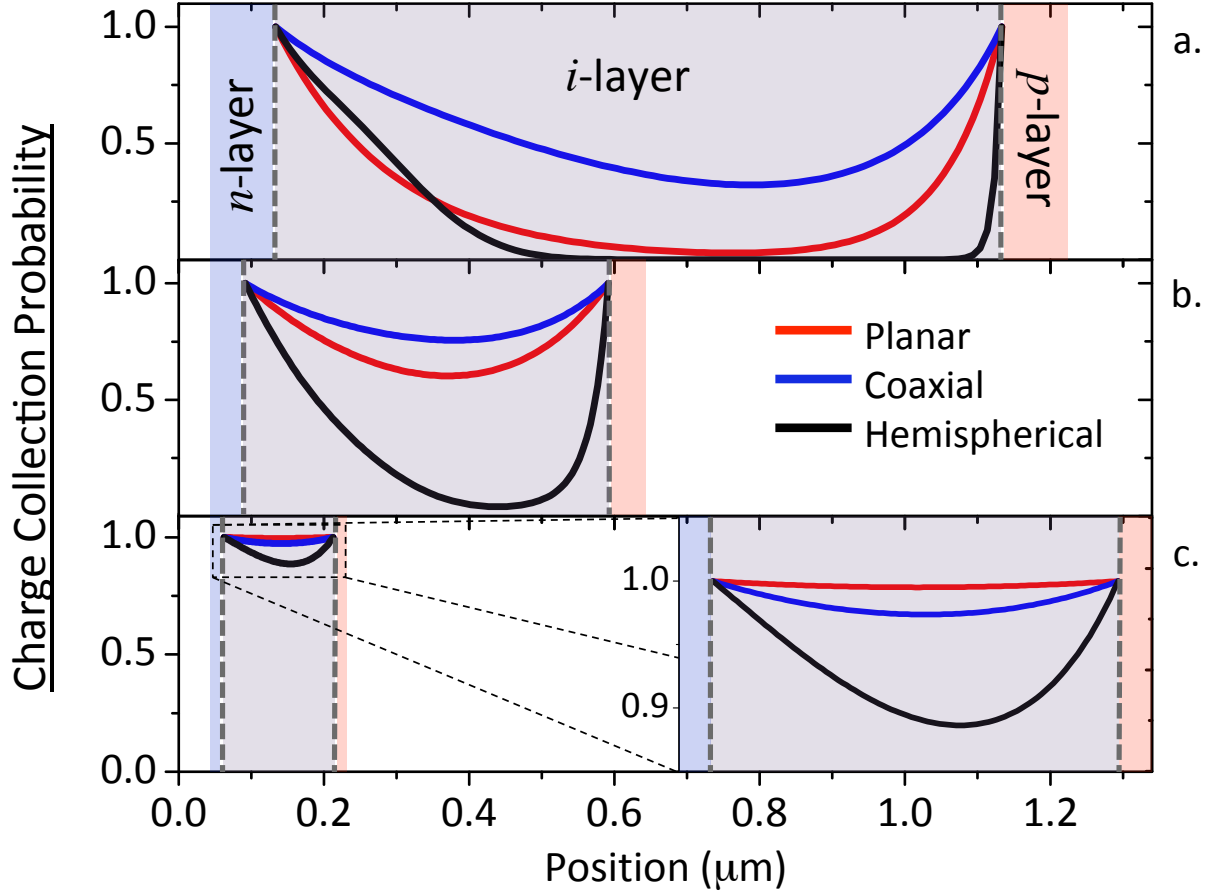

**Fig. 3.** Charge carrier collection probabilities within the *i*-layer for planar (red), coaxial (blue), and hemispherical (black) a-Si:H solar cells, for *i*-layer thicknesses of a) 1000 nm b) 500 nm, and c) 150 nm.

Using ellipsometery data to define the real $n(\varepsilon_\gamma)$ and imaginary $k(\varepsilon_\gamma)$ indices of refraction for a-Si:H [27], we calculate an absorption coefficient $\alpha(\varepsilon_\gamma)$ via the relationship $\alpha(\varepsilon_\gamma) = \frac{4\pi k(\varepsilon_\gamma)}{\lambda(\varepsilon_\gamma)}$, which is used in the equations in Tables 1&2 to calculate the light-current produced for each structure. The 2D hcp lattice for the non-planar structures, which also affects the light-current produced, is shown in Fig. 4. From observation of the 2D hcp lattice, it is seen that some light incident on the array will be lost in the empty space between adjacent cells. The percentage of available area for photovoltaic conversion will vary as a function of the radial size of individual cells. That is, for increasingly thinner *i*-layers in the coaxial and hemispherical architectures, the percentage of empty space in between individual cells decreases.

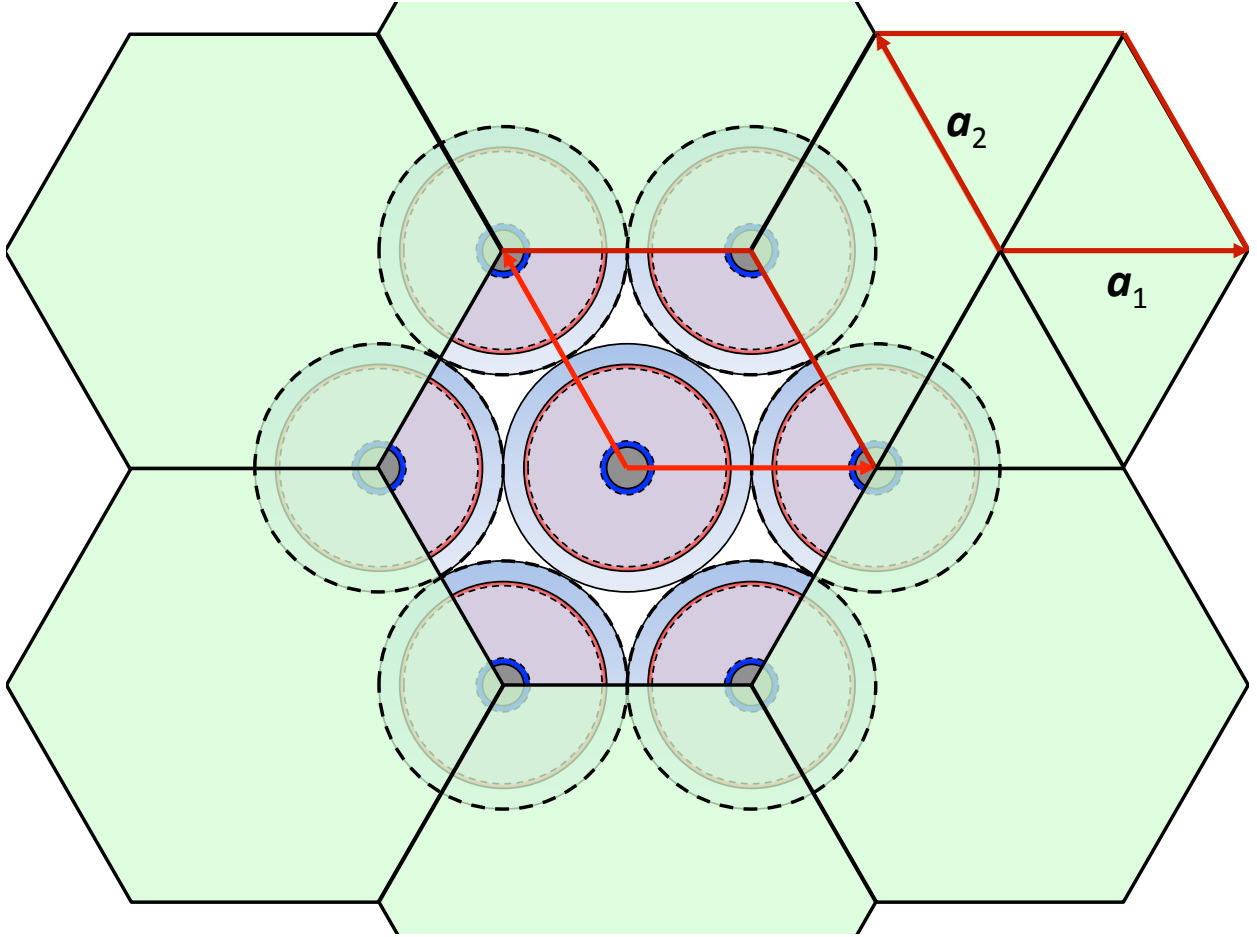

**Fig. 4**. 2D hcp array of nanoscopic coaxial/hemispherical solar cells.

In these simulations, we also consider performance under conditions of light concentration for the non-planar structures. Concern for simulations incorporating light concentration stem from the fact that when spacing between adjacent non-planar cells has sub-wavelength dimensions, the inner metallic contacts can act as optical antennae [28-30], thereby harvesting more light, and consequently increasing short-circuit current produced. We consider two limits; when there is no light concentration, and thus any incident light on the empty space between individual cells is lost, and perfect light concentration, when all incident light on the array is focused into the active photovoltaic area. This will, therefore, yield upper and lower limits for performance of such arrays. We treat the case of perfect light concentration by conserving incident photon flux $\frac{I_{AM1.5}}{\varepsilon_\gamma}$ on the primitive cell into the annulus of individual non-planar cells. Doing so causes photon flux intensity in the annulus to be larger than the actual incident photon flux intensity, since the area of the annulus is smaller than the area of the primitive cell. The increased intensity of the spectral irradiance in the annulus is given by

$$I_{AM1.5+}(\varepsilon_\gamma) = I_{AM1.5}(\varepsilon_\gamma) \frac{2\,r_4^2\sqrt{3}}{\pi\,[r_3^2 - r_0^2]},$$

where $I_{AM1.5}$ is the solar spectrum incident onto the primitive cell, $I_{AM1.5+}$ is the solar spectrum that is concentrated into the annulus, the area of the primitive cell is $2\,r_4^2\sqrt{3}$, and the area of the annulus of the non-planar cell is $\pi [r_3^2 - r_0^2]$. This concentrated spectrum replaces the AM1.5 spectrum in Table 1 when simulating performance under perfect light concentration conditions. The number $N$ of cells in the array is determined from the primitive cell area by $N = \frac{A_{PV}}{2 \cdot 3^{1/2} \cdot r_4^2}$, where, again, $A_{PV} = 1\ cm^2$ for all simulations. Under light concentrating conditions, the performance of a non-planar array behaves as though the number of cells in the array were increased. This can be seen by multiplying the concentrated solar spectrum with the number of cells in the array; *i.e.* $N\ I_{AM1.5+} = \frac{A_{PV}}{\pi [r_3^2 - r_0^2]} I_{AM1.5}$. Hence, the multiplication of the concentrated spectrum and number of cells in the array can be equally described as an effective number of cells in the array $N_{eff} = \frac{A_{PV}}{\pi [r_3^2 - r_0^2]}$ multiplied by the normal AM1.5 solar spectrum; *i.e.* $N\ I_{AM1.5+} = N_{eff}\ I_{AM1.5}$.

By varying coaxial length in the simulations, optimal coaxial performance occurs near a length of 10 μm, under both normal and light concentrating conditions. Fig. 5 shows all three structures plotted versus *i*-layer thickness, with the coaxial plots having lengths of 10 μm. Spanning all *i*-layer thicknesses, the coaxial structure outperforms both the planar and the hemispherical architectures, with a peak efficiency lying at just over 12%. However, the coaxial architecture only outperforms the planar architecture when including perfect light concentration into the model. Because of light concentrating affects, the coaxial structure is most efficient when *i*-layer thickness is small (the efficiency plateaus for *i*-layer thicknesses $\leq 100\ nm$), when charge carrier extraction is maximized. When the coaxial array has no light concentrating effects, it performs similar to the planar device. Experimentally, it could be expected that the coaxial array would perform somewhere between the two limits of concentration considered here, since perfect light concentration is unrealistic in a real device. Based on values for maximum efficiency calculated in these simulations, geometrical variations of the energy band diagram in the SCR do not appear to be affecting performance significantly. For all simulations performed, with and without light concentration, optimal *i*-layer thicknesses occur near or below values where charge carrier collection begins to degrade (compare collection probabilities with *i*-layer thicknesses in Fig. 3). Therefore, while geometrical changes to the energy band diagram most certainly affect charge transport in the SCR, it is the effect that the sub-wavelength array has on light harvesting that seems to be most significantly impacting performance from one architecture to another.

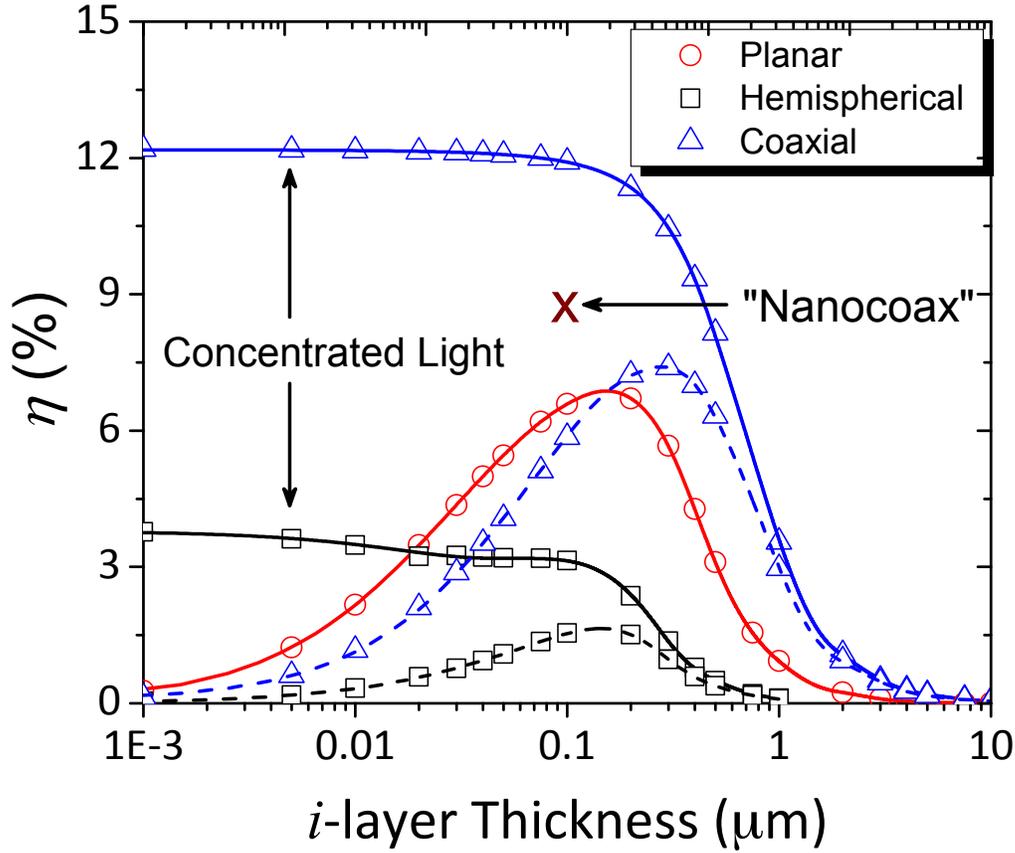

**Fig. 5**. Efficiency vs. *i*-layer thickness curves. The dashed lines represent the efficiency curves when less than 100% light absorption occurs for the coaxial and hemispherical structures. The "x" marks the approximate location of the efficiency observed experimentally for the nanocoax in 2010.

*I-V* curves representative of maximum efficiency in each geometry are shown in Fig. 6. The *i*-layer thicknesses used are 150 nm for the planar, 5 nm for the coaxial, and 5 nm for the hemispherical, as these are the approximate optimal spatial parameters that maximize efficiency in each structure. Comparison between experimentally-fabricated planar and nanocoaxial a-Si:H solar cells was reported in 2010 [24]. In this study, it was observed that efficiency improvement for the nanocoaxial array arose from enhanced absorption properties associated with these devices; even greater than that observed for texturized and back reflector superstrate a-Si:H solar cells [24]. For the planar a-Si:H solar cell reported in that study, the short-circuit current value obtained is nearly identical to the value we calculate using our simulation (see Fig. 6). We note that in [24], the *i*-layer thickness of the planar cell was 90 nm, while the *I-V* trace we refer to in Fig. 6 has an *i*-layer thickness of 150 nm. However, according to Fig. 5, performance of the planar architecture will not change appreciably between *i*-layer thicknesses of 90 to 200 nm, due to the breadth of the efficiency peak; hence short-circuit current for the planar architecture in our simulation will be nearly identical when *i*-layer thickness is 90 nm. Similarly, short-circuit current observed for the nanocoaxial array very closely resembles the maximum short-circuit current observed in our simulations. Again, an important distinction between the two is that *i*-layer thickness and length of the nanocoax in [24] were 90 nm and 1.5 $\mu$m, respectively, while the *i*-layer thickness and length of the nanocoaxial array in Fig. 6 is 5 nm and 10 $\mu$m, respectively. As before with the comparison between the planar cell in [24] and our simulation of the planar architecture, based on Fig. 5, our calculated efficiency for the nanocoaxial array does not change appreciably when *i*-layer thickness is less than 100 nm. As such, the efficiency we calculate when *i*-layer thickness is 5 nm will be nearly identical when *i*-layer thickness is 90 nm. For all simulations performed, the open-circuit voltage for each structure exceeds those reported in [24]. We accredit this to other forms of recombination not included in the recombination rate; namely band tail and surface recombination. Excluding band tail recombination, the open-circuit voltages obtained seem perfectly acceptable compared to experimental values. The efficiency reported for the nanocoax in [24] is approximately 8%, which lies directly between the two limits of concentration considered in our simulations (see Fig. 5). Therefore, based on results for planar and nanocoax solar cells reported in [24], we conclude that our simulation results closely match those experiments and give insight into the nature of increased absorption observed experimentally for the nanocoax array.

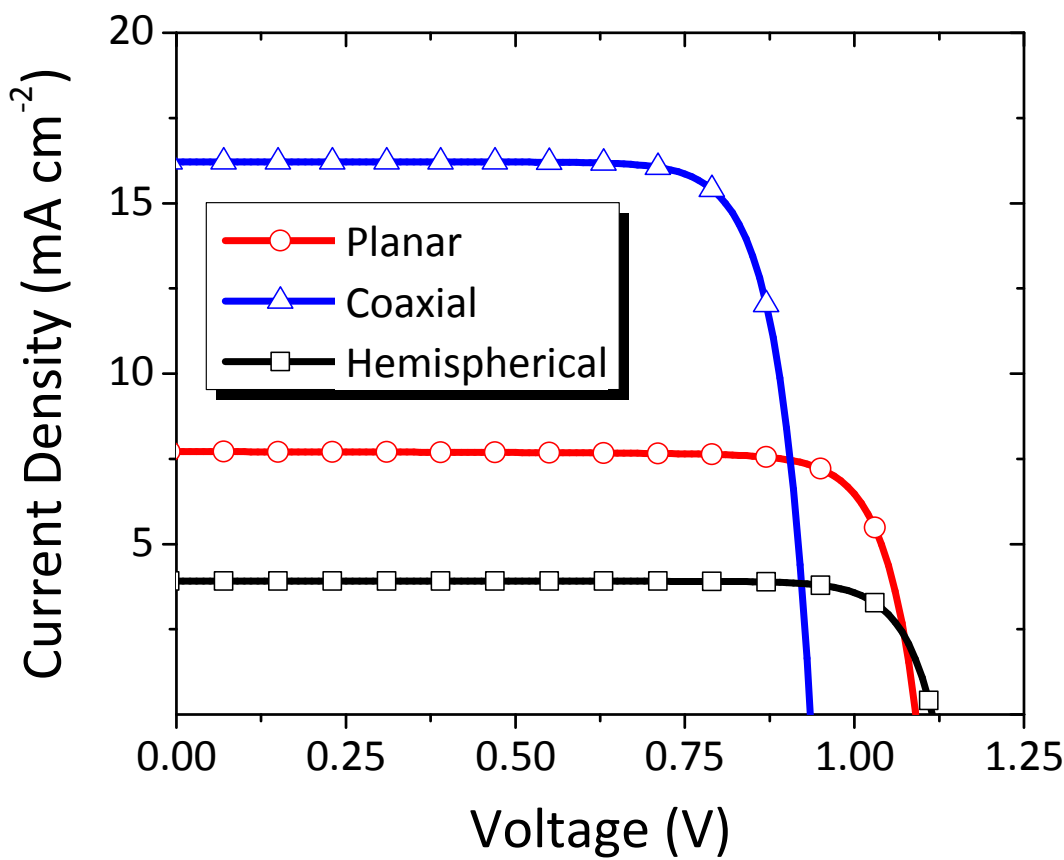

**Fig. 6**. *I-V* curves for optimized spatial parameters within each geometry, performed over an area of 1 cm². The *i*-layer thicknesses are 150 nm for the planar, 5 nm for the coaxial, and 5 nm for the hemispherical structures. The coaxial structure is 10 μm long. The *I-V* curves shown represent current under 100% light absorption.

The simulations performed herein utilize an analytical model [23] describing the geometrically generalized device physics of *p-n* junctions. However, due to the nature of the *n-i-p* stacking of a-Si:H solar cells, only a subsection of that model was used here; *i.e.* the generalized description in the SCR. For thin film solar cells in other material systems (*e.g.* CdTe, CIGS, *etc*.), which are not dominated by charge drift in the SCR, it will be necessary to include diffusional charge transport from the doped quasi-neutral regions to accurately model their performance. While not discussed in this work, solutions for diffusional charge transport will entail numerically solving non-linear, inhomogeneous, partial differential equations when working in cylindrically or spherically symmetric geometries. By combining the solutions for diffusional transport from the quasi-neutral regions with the drift transport from the SCR, a *p-n* junction solar cell constructed in any symmetrical geometry, in any material system, can be mathematically modeled using this generalized formalism.

## 4. Section IV (Conclusion)

We have simulated performance for a-Si:H *n-i-p* solar cells in planar, coaxial, and hemispherical architectures. Our simulations for the planar structure indicate that the *p*- and *n*-layers negligibly impact device performance. In addition, our simulations for the planar geometry are in good agreement with simulations performed using SCAPS-1D. Efficiency improvements for a-Si:H solar cells are possible for the coaxial architecture only when taking light concentrating effects into account. These improvements appear to stem from the fact that surface-area and cell-number are maximized for the coaxial array under the light concentrating conditions, and not from changes in the built-in electric field as a function of geometry. Performance of planar and nanocoaxial a-Si:H solar cells calculated in these simulations closely resemble those reported experimentally in 2010 and give insight into the nature of increased absorption observed experimentally for the nanocoax array.

*Acknowledgments*

The authors would like to thank Dr. Constantin Andronache at Boston College for useful conversations regarding the setup of these simulations. The authors would also like to thank Prof. Nikolas Podraza, University of Toledo, for performing ellipsometry on films grown at Solasta Inc.

## 5. Section VI (References)

**Appendix**. Simulation parameters and values.

| Symbol | Values | |
|---|---|---|
| $q$ | Fundamental unit of charge | $1.602 \times 10^{-19}$ [C] |
| $\hbar$ | Planck's constant | $1.055 \times 10^{-34}$ [J s] |
| $k_B$ | Boltzmann's constant | $1.38 \times 10^{-23}$ [J $K^{-1}$] |
| c | Speed of light in vacuum | $3.0 \times 10^{10}$ [cm $s^{-1}$] |
| $T_A$ | Ambient temperature of solar cell | 300 [K] |
| $\beta_A$ | Inverse thermal energy of ambient temperature | $(k_B T_A)^{-1}$ [$J^{-1}$] |
| $N_C$ | Conduction band effective density of states | $2.5 \times 10^{20}$ [$cm^{-3}$] |
| $N_V$ | Valence band effective density of states | $2.5 \times 10^{20}$ [$cm^{-3}$] |
| $N_D$ | Concentration of donor atoms/free electrons in n-type region | $8.0 \times 10^{18}$ [$cm^{-3}$] |
| $N_A$ | Concentration of acceptor atoms/free holes in p-type region | $3.0 \times 10^{18}$ [$cm^{-3}$] |
| $\Delta_{SC}$ | Band gap of intrinsic amorphous silicon | 1.8 [eV] |
| $n_i$ | Intrinsic charge carrier concentration in $i$-layer | $\sqrt{N_C N_V} \, exp\left(-\beta_A \frac{\Delta_{SC}}{2}\right)$ [$cm^{-3}$] |
| $t_M$ | Thickness of TCO window | $50.0 \times 10^{-7}$ [cm] |
| $r_0$ | Thickness of back contact/origin offset | $50.0 \times 10^{-7}$ [cm] |
| $t_{SC}$ | Thickness of $i$-layer/space-charge region | Batching parameter [cm] |
| $t_N(t_{SC})$ | Thickness of n-type region | $\frac{t_{SC}}{12}$ [cm] |
| $t_P(t_{SC})$ | Thickness of p-type region | $\frac{t_{SC}}{10}$ [cm] |
| $r_1(t_{SC})$ | $n$-type region edge | $r_0 + t_N(t_{SC})$ [cm] |
| $r_2(t_{SC})$ | Space-charge region edge | $r_1(t_{SC}) + t_{SC}$ [cm] |
| $r_3(t_{SC})$ | $p$-type region edge | $r_2(t_{SC}) + t_P(t_{SC})$ [cm] |
| $r_4(t_{SC})$ | Front surface of cell | $r_3(t_{SC}) + t_M$ [cm] |
| $\varepsilon_\gamma$ | Photon energy | Integration variable [J] |
| $\lambda(\varepsilon_\gamma)$ | Photon wavelength | $\frac{2\pi\hbar c}{\varepsilon_\gamma}$ [cm] |
| $A_{PV}$ | Area of solar cell | 1.0 [cm$^2$] |
| $N_1(t_{SC})$ | Number of solar cells | $\frac{A_{PV}}{2\, r_4(t_{SC})^2\, \sqrt{3}}$ |
| $N_2(t_{SC})$ | Effective number of solar cells under light concentration | $\frac{A_{PV}}{\pi\, [r_3(t_{SC})^2 - r_0^2]}$ |
| $\chi_{SC}$ | Electron affinity in *i*-layer | 3.9 [eV] |
| $V$ | Applied bias | Independent variable [V] |
| $\varepsilon_{max}(V)$ | Maximum absorbed photon energy | $\chi_{SC} + \Delta_{SC} - V$ [eV] |
| $\mu_n$ | Electron mobility in *i*-layer | 1.0 [cm$^2$ $V^{-1}s^{-1}$] |
| $\mu_p$ | Hole mobility in *i*-layer | 0.01 [cm$^2$ $V^{-1}s^{-1}$] |
| $\tau_n$ | Electron lifetime in *i*-layer | 1.0×10$^{-9}$ [s] |
| $\tau_p$ | Hole lifetime in *i*-layer | 5.0×10$^{-9}$ [s] |
| $\Lambda_n$ | Electron Auger recombination coefficient in *i*-layer | 0.3×10$^{-30}$ [cm$^6 s^{-1}$] |
| $\Lambda_p$ | Hole Auger recombination coefficient in *i*-layer | 1.1×10$^{-30}$ [cm$^6 s^{-1}$] |
| $B$ | Radiative recombination coefficient | 1.1×10$^{-14}$ [cm$^3 s^{-1}$] |
| $V_{BI}$ | Built-in junction bias | $\frac{1}{q\,\beta_A} ln\left(\frac{N_D N_A}{n_i^2}\right)$ [V] |